\documentclass[twocolumn]{aastex631}

\usepackage{newtxtext,newtxmath}
\usepackage{verbatim}

\usepackage[T1]{fontenc}
\usepackage{mathtools}

\DeclareRobustCommand{\VAN}[3]{#2}
\let\VANthebibliography\thebibliography
\def\thebibliography{\DeclareRobustCommand{\VAN}[3]{##3}\VANthebibliography}

\usepackage{soul}
\usepackage{graphicx}	
\usepackage{amsmath}	
\newcommand{\TESS}{\textit{TESS}}
\newcommand{\Kepler}{\textit{Kepler}}
\newcommand{\rearth}{\ensuremath{R_\oplus}}

\newcommand{\Gaia}{\textit{Gaia}}

\usepackage{xcolor}

\definecolor{my_color}{HTML}{3a18b1}

\definecolor{new_color}{HTML}{CF0000}

\shortauthors{Christian et al.}
\shorttitle{Binary-Orbit/Planet-Orbit Alignment}

\begin{document}

\title{Wide Binary Orbits are Preferentially Aligned with the Orbits of Small Planets, but Probably Not Hot Jupiters}

\correspondingauthor{Sam Christian}
\email{samchristian@mit.edu}

\author[0000-0003-0046-2494]{Sam Christian}
\affiliation{Department of Physics and Kavli Institute for Astrophysics and Space Research, Massachusetts Institute of Technology, Cambridge, MA 02139, USA}
\author[0000-0001-7246-5438]{Andrew Vanderburg}
\affiliation{Department of Physics and Kavli Institute for Astrophysics and Space Research, Massachusetts Institute of Technology, Cambridge, MA 02139, USA}

\author[0000-0002-7733-4522]{Juliette Becker}
\affiliation{Department of Astronomy, University of Wisconsin-Madison, Madison, WI 53706, USA}
\author[0000-0001-9811-568X]{Adam L. Kraus}
\affiliation{Department of Astronomy, University of Texas at Austin, Austin, Texas 78712, USA}
\author[0000-0003-3904-7378]{Logan Pearce}
\affiliation{Steward Observatory, University of Arizona, Tucson, AZ 85721, USA}
\author[0000-0001-6588-9574]{Karen A.\ Collins}
\affiliation{Center for Astrophysics \textbar \ Harvard \& Smithsonian, 60 Garden Street, Cambridge, MA 02138, USA}

\author{Malena Rice}
\affiliation{Department of Astronomy, Yale University, New Haven, CT 06511, USA}

\author[0000-0002-4625-7333]{Eric L.\ N.\ Jensen}
\affiliation{Department of Physics \& Astronomy, Swarthmore College, Swarthmore PA 19081, USA}

\author[0000-0002-2970-0532]{David Baker} 
\affiliation{Physics Department, Austin College, Sherman, TX 75090, USA}
\author{Paul Benni}
\affiliation{Acton Sky Portal (private observatory), Acton, MA USA}
\author[0000-0001-6637-5401]{Allyson Bieryla} 
\affiliation{Center for Astrophysics \textbar \ Harvard \& Smithsonian, 60 Garden Street, Cambridge, MA 02138, USA}
\author{Avraham Binnenfeld}
\affiliation{Porter School of the Environment and Earth Sciences, Raymond and Beverly Sackler Faculty of Exact Sciences, Tel Aviv University, Tel Aviv, 6997801, Israel}
\author[0000-0003-2781-3207]{Kevin I.\ Collins}
\affiliation{George Mason University, 4400 University Drive, Fairfax, VA, 22030 USA}
\author[0000-0003-2239-0567]{Dennis M.\ Conti}
\affiliation{American Association of Variable Star Observers, 185 Alewife Brook Parkway, Suite 410, Cambridge, MA 02138, USA}
\author{Phil Evans}
\affiliation{El Sauce Observatory, Coquimbo Province, Chile}
\author{Eric Girardin}
\affiliation{Grand Pra Observatory, Switzerland}
\author{Joao Gregorio}
\affiliation{Crow Observatory, Portalegre, Portugal}
\author{Tsevi Mazeh} \affiliation{School of Physics and Astronomy, Tel Aviv University, Tel Aviv, 6997801, Israel}
\author{Felipe Murgas}
\affiliation{Instituto de Astrof\'isica de Canarias (IAC), E-38205 La Laguna, Tenerife, Spain}
\affiliation{Departamento de Astrof\'isica, Universidad de La Laguna (ULL), E-38206 La Laguna, Tenerife, Spain}
\author[0000-0001-5850-4373]{Aviad Panahi} \affiliation{School of Physics and Astronomy, Tel Aviv University, Tel Aviv, 6997801, Israel}
\author[0000-0003-1572-7707]{Francisco J. Pozuelos} 
\affiliation{Space Sciences, Technologies and Astrophysics Research (STAR) Institute, Université de Liège, 19C Allée du 6 Août, 4000 Liège, Belgium} 
\affiliation{Astrobiology Research Unit, Université de Liège, 19C Allée du 6 Août, 4000 Liège, Belgium}
\author{Howard M. Relles}
\affiliation{Center for Astrophysics \textbar \ Harvard \& Smithsonian, 60 Garden Street, Cambridge, MA 02138, USA}
\author{Fabian Rodriguez Frustaglia}
\affiliation{American Association of Variable Star Observers, 49 Bay State Road, Cambridge, MA 02138, USA}
\author[0000-0001-8227-1020]{Richard P.\ Schwarz}
\affiliation{Center for Astrophysics \textbar \ Harvard \& Smithsonian, 60 Garden Street, Cambridge, MA 02138, USA}
\author{Gregor Srdoc}
\affiliation{Kotizarovci Observatory, Sarsoni 90, 51216 Viskovo, Croatia}
\author[0000-0003-2163-1437]{Chris Stockdale}
\affiliation{Hazelwood Observatory, Australia}
\author[0000-0001-5603-6895]{Thiam-Guan Tan}
\affiliation{Perth Exoplanet Survey Telescope, Perth, Western Australia}
\affiliation{Curtin Institute of Radio Astronomy, Curtin University, Bentley, Western Australia 6102}
\author[0000-0002-8961-0352]{William~C.~Waalkes}
\affiliation{Department of Physics and Astronomy, Dartmouth College, Hanover NH 03755, USA}
\author[0000-0003-3092-4418]{Gavin Wang}
\affiliation{Department of Physics \& Astronomy, Johns Hopkins University, 3400 N. Charles Street, Baltimore, MD 21218, USA}
\author[0000-0002-7424-9891]{Justin Wittrock}
\affiliation{George Mason University, 4400 University Drive, Fairfax, VA, 22030 USA}
\author{Shay Zucker}
\affiliation{Porter School of the Environment and Earth Sciences, Tel Aviv University, Tel Aviv 6997801, Israel}

\begin{abstract}
Studying the relative orientations of the orbits of exoplanets and wide-orbiting binary companions (semimajor axis greater than 100 AU) can shed light on how planets form and evolve in binary systems. Previous observations by multiple groups discovered a possible alignment between the orbits of visual binaries and the exoplanets that reside in them. In this study, using data from \textit{Gaia} DR3 and TESS, we confirm the existence of an alignment between the orbits of small planets $(R<6 R_\oplus)$ and binary systems with semimajor axes below 700 AU ($p=10^{-6}$). However, we find no statistical evidence for alignment between planet and binary orbits for binary semimajor axes greater than 700 AU, and no evidence for alignment of large, closely-orbiting planets (mostly hot Jupiters) and binaries at any separation. The lack of orbital alignment between our large planet sample and their binary companions appears significantly different from our small planet sample, even taking into account selection effects. Therefore, we conclude that any alignment between wide-binaries and our sample of large planets (predominantly hot Jupiters) is probably not as strong as what we observe for small planets in binaries with semimajor axes less than 700 AU. The difference in the alignment distribution of hot Jupiters and smaller planets may be attributed to the unique evolutionary mechanisms occuring in systems that form hot Jupiters, including potentially destabilizing secular resonances that onset as the protoplanetary disk dissipates and high-eccentricity migration occurring after the disk is gone. 
\end{abstract}

\keywords{planet-star interactions}


\section{Introduction}
Most stars in our galaxy reside in binary or higher-order stellar systems \citep{Raghavan2010ApJS}, and many of these systems are known to host exoplanets \citep{Mugrauer2019}. Using a variety of methods, astronomers have detected exoplanets on a range of different orbits in binary systems, which can generally be categorized into two main configurations: circumbinary, where the planet orbits both stars in the binary system \citep[e.g.][]{Doyle2011Sci, Orosz2012Sci}, and circumstellar, where the planet orbits only one host star \citep[e.g.][]{Anglada-Escude2016Natur, Mugrauer2019}. In this paper, we focus on exoplanets in circumstellar configurations in wide binary systems, which happens to be the most common scenario for known exoplanets \citep{Mugrauer2019}. For the purposes of this paper, we define a wide binary to be any binary star system with semimajor axis $\gtrapprox 100$ AU.

The effect of wide-binaries on the orbits of exoplanets in those systems has been the subject of intense study for years. Generally, it is understood that gravitational interactions with wide stellar binary companions can influence the orbits of planets on long timescales. For example, the Lidov-Kozai-Von-Zeipel (LKZ) mechanism \citep{Vonzeipel1910,lidov,kozai}, induced by the binary companion, could provide an explanation for at least some of the orbits of hot Jupiters \citep{Wu2003,Fabrycky2007,Petrovich2015b,Dawson2018}, although other mechanisms that can explain the existence of hot Jupiters exist, and it is not possible for the LKZ effect to explain the existence of all Hot Jupiters \citep{Becker2015ApJL, Ngo2016}. Moreover, occurrence rate studies from the Kepler mission have shown that the presence of binary companions with separations less than about 50 astronomical units can suppress planet formation \citep{Kraus2016AJ, Moe2021MNRAS}, presumably due to disruption of the protoplanetary disk. And finally, theoretical work \citep{Batygin2012Natur,Lai2014,ZanazziLai2018MNRAS} (and tentative observations, \citealt{JensenAkeson2020, Hjorth2021PNAS,Offner2023}) has suggested that binary companions can align protoplanetary disks with the binary's orbital plane, giving rise to systems with angular momenta well aligned between planets and binary stars, yet misaligned with the stellar spin axis.  

Recently, several studies have focused on exploring this phenomenon in more detail, and found evidence suggesting that the orbits of planetary systems and the binaries they reside in are aligned\footnote{ It is important to distinguish orbit-orbit alignment from spin-orbit alignment. Spin-orbit alignment is the alignment between the stellar spin axis and exoplanet orbit, \citep[e.g.][]{Albrecht2022}. Orbit-orbit alignment is an alignment between the exoplanet orbit and binary system orbit.} \citep{Behmard2022,Christian2022,Dupuy2022}.  \citet{Christian2022} and \citet{Behmard2022} independently found an alignment in \textit{TESS} exoplanets and visual binaries identified from \textit{Gaia} DR2 data extending from 100 to over $10^4$ AU. Meanwhile \citet{Dupuy2022}, and more recently \citet{Lester2023AJ}, found a similar pattern through astrometric monitoring of binary systems hosting planets discovered by the \Kepler\ mission. Thanks to the high resolution adaptive optics and speckle imaging observations, these binaries could be resolved at closer separations than \citet{Christian2022} and \citet{Behmard2022}. These works all concluded that closely orbiting planets and wide binaries show at least some preferential alignment, and \citet{Behmard2022}'s results hinted at a dichotomy between the orbit/orbit alignment states for small planets and hot Jupiters.

These studies were limited in the conclusions they could draw, primarily because any measurement of transiting planet orbit/binary orbit misalignment must be statistical in nature. In particular, this is because for transiting planet systems, it is only possible to measure the relative \textit{inclination} difference between binary and exoplanet orbit. Because inclinations are measured relative to our line of sight, the difference in inclination between the planet and stellar orbits is equal to the minimum misalignment between orbits, so measuring an inclination difference of zero degrees, for instance, could indicate a completely aligned or completely misaligned system. But by measuring the inclination differences for a large sample of systems, a preponderance of systems with small inclination differences implies a preponderance of aligned systems. As a result of the ambiguity, large samples are critical for confidently assessing the presence of aligned populations and identifying any correlations with alignment or misalignment.  

In this work, we therefore expand upon earlier works examining orbit/orbit alignment with a significantly larger sample of planetary systems than were previously studied. In particular, we perform a similar analysis to that of \citet{Christian2022} and constrain the orbital inclination of a large sample of wide binary star orbits in systems that have transiting planet candidates from the TESS mission. Thanks to several years of new planet candidate discoveries from TESS, new wide binaries detected in Gaia DR3, and more precise astrometric solutions, we are able to draw stronger and more nuanced conclusions on planet/binary alignment. While we find strong evidence for alignment between small planets $(R<6 R_\oplus)$ and relatively close binaries ($a<700$AU), we find no compelling evidence for alignment in systems with giant planets or binary separations greater than $a\gtrsim700$AU. 

Our paper is organized as follows. We describe the construction of our samples (both a sample of transiting planetary systems with wide binary companions, and a control sample of wide binaries without any preference for transiting planets) in Section \ref{sec:Observations} and describe our analysis of the data in Section \ref{sec:analysis}. We report our major results in Section \ref{sec:results} and discuss how they compare with previous studies and what these trends imply for the early dynamical history of these systems in Section \ref{sec:discussion}. We then conclude in Section \ref{sec:conclusions} with a summary of our results and the most significant implications. 





\section{Observations/Data}
\label{sec:Observations}

\subsection{Selection of wide binaries}
\label{sec:selection}
We start by gathering the sample of wide binary stars identified using Gaia EDR3 data by \citet{elbadry2021}. This updated sample contains approximately one million common-proper-motion wide binary pairs within about one kiloparsec of the Sun. This work builds upon a previous catalog compiled by \citet{elbadry2018} (and used by \citealt{Christian2022}) using data from Gaia DR2, which focused on systems within 200 pc, and included only about 50,000 binaries. Like \citet{Christian2022}, we removed systems where the astrometric solutions showed evidence of excess scatter, in particular by removing systems where at least one star -- either primary or companion -- has a Renormalized Unit Weight Error (RUWE) greater than 1.4, following the recommendations of \citet{lindegren2018}. We also removed putative binary systems that might instead be chance alignments. We used the pseudo-probability of chance alignments, $R$, as reported by \citet{elbadry2021}, to identify these systems. We impose a relatively stringent requirement that the binary systems have $R$ less than 0.01 to remain in our sample to ensure high purity. 

We removed any systems with white dwarf companions. Post main sequence evolution likely changes the architecture of the binary systems significantly \citep{Veras2016}, and might affect the resulting inclination distribution. Using only main sequence and red giant stars allows us to study potential primordial or early-lifetime mechanisms that contributed to the observed alignment distribution.




\subsection{Selection of exoplanets}

Like \citet{Christian2022}, we source transiting exoplanet candidates from the TESS Object of Interest catalog \citep{guerrero}, which we downloaded on 1 June 2022. The TESS exoplanets are listed by their identifiers in the the TESS Input Catalog (TIC), which is based on \textit{Gaia} DR2. We crossmatched the \Gaia\ DR2-based TESS planet candidates with the \Gaia\ DR3-based  \citet{elbadry2021} sample of wide binaries using the \textit{Gaia} archive's \textit{Gaia} DR2 to \textit{Gaia} DR3 crossmatch.

After identifying the transiting planet candidates in binary systems via this crossmatch, we excluded known and suspected false positives from our sample. In particular, we performed the following steps: 

\begin{enumerate}
    \item We queried the status of the planet candidates on ExoFOP-TESS to identify which have been shown to be false positives or false alarms. These are confidently known false positives, based on analysis of TESS data performed after the planet candidates were initially identified, or follow-up observations (mostly photometric or spectroscopic). 
    \item Beyond the planet candidates that were identified as confident false positives (FPs) on ExoFOP-TESS, we removed suspected FPs based on photometric follow-up observations by Sub-group 1 (SG1) the TESS Follow-up Working Group (TFOP WG), following \citet{Christian2022}, Section 2.1.3. 
\end{enumerate}

The resulting sample contains 398 exoplanet candidates in binary systems (in addition to the previously described cuts on binary RUWE and false positive probability as described in section \ref{sec:selection}). This is a significantly larger sample than the 67 objects used by \citet{Christian2022}. Of these 398 systems, 59 are known exoplanet systems discovered before the \TESS \,mission, and 33 were confirmed to be exoplanets by the \TESS\, mission. As of April 2024, 150 are designated by SG1 as verified planet candidates, verified candidates with possible contaminated aperture, or cleared planet candidates, indicating that most potential sources of a false-positive detection have been ruled out, but not to a degree to designate the planet as confirmed.

Overall, only 79 systems in our sample that were not previously known to be confirmed planets lack significant follow-up observations from TFOP SG1.



\begin{figure*}
    \includegraphics[width=.9\textwidth]{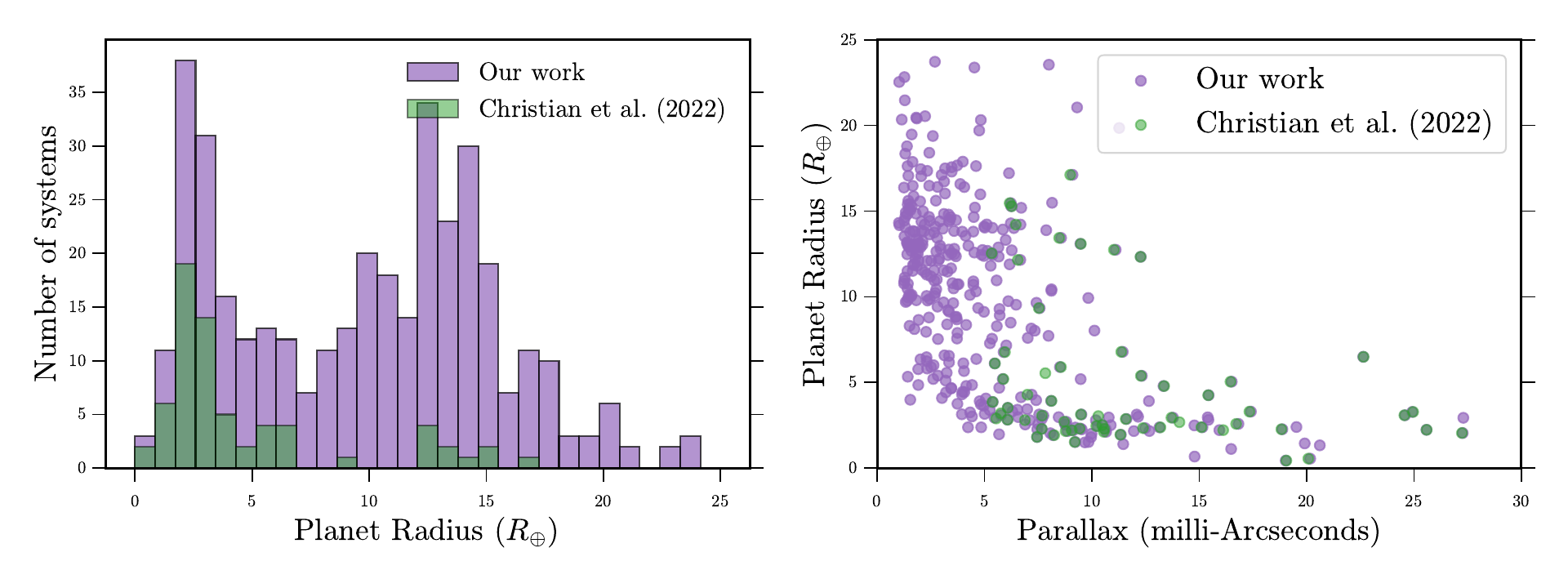}
    \caption{Comparisons of planet radius and parallax distributions from our sample to that of \citet{Christian2022}. \textit{Left:} a comparison of the histograms of planet radius in the new and old samples. \textit{Right:} a comparison of the joint distribution of the parallax and planet size between the sample in this work and that of \citet{Christian2022}.  The significant increase in new giant planets in our sample is primarily due to the \citet{elbadry2021} catalog extending to further distances (smaller parallaxes) than that of \citet{elbadry2018}. Large transiting planets are inherently rarer, but easier to detect around more distant/faint stars. In the right-hand panel, the points corresponding to the same systems in the old and new samples do not precisely line up because of the difference of \textit{Gaia} DR2 and DR3 parallaxes and changes in the \TESS\ planetary radii as new data have been collected. }
    \label{fig:previousWork}
\end{figure*}

The increase in number of systems in our sample is accompanied by changes in the characteristics of the typical system in our sample compared to that of \citet{Christian2022}. Figure \ref{fig:previousWork} shows histograms that illustrate these changes. Compared to the sample of \citet{Christian2022}, ours is a) more dominated by giant planets, especially hot Jupiters, and b) composed of more distant stars. These changes are explained by the fact that the \citet{elbadry2021} catalog extends to greater distances than previous work, thereby increasing the number of faint binary systems. Moreover, the significant increases in the TOI catalog are largely due to the TESS faint star search \citep{Kunimoto2022ApJS}, which increases the magnitude limit to which planet candidates are reported from the MIT Quick Look Pipeline from 10.5 to 13.5 in TESS-band magnitudes. Both of these changes increased the number of faint, distant stars in our sample, and at these faint magnitudes, TESS is primarily sensitive to short-period gas giant exoplanets. As a result, our sample has a much greater proportion of hot Jupiter exoplanets than that of \citet{Christian2022}.

\subsection{Control Sample}

In order to control for unknown biases in the inclination distribution of binaries from \textit{Gaia} DR3, we build a control sample of systems without detected exoplanets. As in \citet{Christian2022}, we selected the members of our control sample by searching for ``similar'' star systems to our exoplanet hosts. We quantified the similarity between our planet-hosting systems and putative members of our control sample using a slightly different metric $\mathcal{M}$ than \citet{Christian2022}. In particular, we defined:
\begin{equation}
    \mathcal{M}=(\Delta G_1)^2+(\Delta G_2)^2+\left(\frac{\Delta \varpi_1}{\varpi_{1,\rm exo}}\right)^2+\left(\frac{\Delta s}{s_{\rm exo}}\right)^2
\end{equation}

\noindent where $G_1$ and $G_2$ are the \textit{Gaia} G-band magnitudes of the primary and secondary stars, $\varpi_1$ is parallax of the primary stars in each binary system, and $s$ is binary angular separation. Here, we use $\Delta$ to refer to the difference in each quantity between that of the exoplanet-host system and the values of putative control-sample members, and the subscript $_{\rm exo}$ denotes the particular value of the exoplanet-hosting system. Matching on apparent magnitude and parallax is a proxy for matching the luminosity of the binary system. There is tentative evidence that the inclination distribution of binary systems changes as a function of mass ratio \citep{Gerbig2024inPrep}, so this is an important quantity to control for. The primary difference between this metric and that used by \citet{Christian2022} is that we do not compute the normalized difference in magnitude, as the magnitude already is defined on a log-scale.

For each binary system in our sample of binaries with exoplanets, we select the ten systems from the \citet{elbadry2021} sample that, after passing the initial quality cuts of RUWE and R described in Section \ref{sec:selection}, have the lowest value of $\mathcal{M}$.


\section{Analysis}
\label{sec:analysis}
\subsection{Stellar Mass Determination}
\label{sec:mass}
\begin{figure}
    \includegraphics[width=\columnwidth]{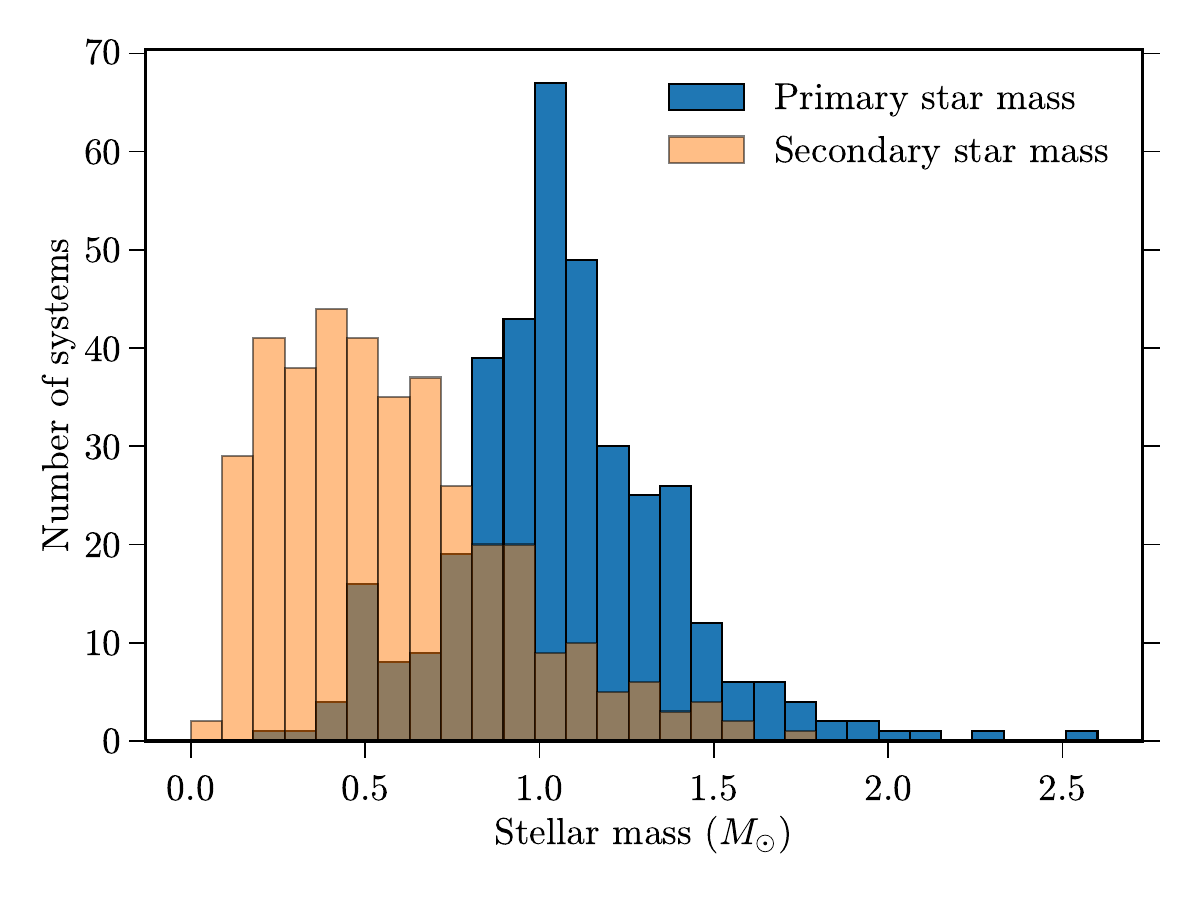}
    \caption{Histogram of the primary companion and secondary companion stellar masses. The primary companion of a binary star system is defined as the star which the exoplanet orbits.}
    \label{fig:massDistribution}
\end{figure}

First, we estimated the masses of each stellar component (primary and secondary) in both our exoplanet and control samples. Like \citet{Christian2022}, we use two different methods: an empirical relation for low-mass M-dwarfs, and fitting to models for higher-mass stars. 

For the low-mass M-dwarf stars -- that is, stars whose absolute $K$-band magnitudes are between 4.5 and 10.5 -- we followed \citet{Christian2022} and estimated masses using the empirical $M_\odot-K$ relationship of \citet{Mann2019}, implemented in an accompanying python package\footnote{\url{https://github.com/awmann/M_-M_K-}}. For the absolute magnitude determination for M-dwarfs, we use the geometric distances of \citet{Bailer2021}.

For the higher-mass stars, we again followed \citet{Christian2022} and estimated masses using the \verb|Isochrones| package \citep{Morton2015}, which infers fundamental stellar parameters using nested sampling to fit stellar evolutionary and atmospheric models given data like broadband apparent magnitudes, spectroscopic metallicity, and parallax. For each star in each of our binary systems, we input the G,BP, and RP \textit{Gaia} magnitudes, as well as J,H, and K magnitudes from the 2MASS survey (if available), and the star's \textit{Gaia} parallax. Unlike \citet{Christian2022}, we include only metallacity estimates from \Gaia, and do not make use of any other spectra (either from ground-based surveys or follow-up observations). \textit{Gaia} DR3 provides low-resolution spectroscopy for certain systems, including derived metallicity. When available, we use the provided spectroscopic metallicity values from \textit{Gaia} DR3 and cut off values below -1 \citep{Fouesneau2022}. We find for the stars in our sample with both spectroscopic and photometric metallicities, the values generally agree, and choose to use the spectroscopic metallicities. 

The mass distribution of the primary and secondary stars is plotted in Figure \ref{fig:massDistribution}. A large fraction of the secondary stars are M-dwarfs, while the primary stars generally have masses above one stellar mass.



\subsection{LOFTI modeling}
Like \citet{Christian2022}, once we had mass estimates for all of the stars in both our exoplanet and control samples, we  we fitted the orbits of the binary systems using Linear Orbits for The Impatient \citep[LOFTI]{Pearce2020}. LOFTI uses a Monte Carlo approach to perform rejection sampling in a simplified 4D orbital parameter space \citep{Blunt2017AJ}. For each system, given the stellar masses, LOFTI randomly generates model orbits, compares them to observational data (i.e. positions, parallax, proper motions, and radial velocities), and either accepts or rejects the samples based on the the goodness of fit. LOFTI repeats this process until a user-specified number of accepted samples is reached; the distribution of the accepted samples then represents the posterior probability distribution for orbital parameters.  



We ran LOFTI on each system in both our exoplanet and control samples. For each system, we input the positions, proper motions, parallax, and radial velocities (where applicable) for each star from Gaia DR3. Because of the significantly larger number of systems in our sample compared to \citet{Christian2022}, we chose to decrease the number of accepted samples per system from 100,000 to 10,000. We found that 10,000 orbits were plenty for us to make reasonable statistical estimates, and this sped up computation by a factor of 10. 

For a small number of systems (3 in our exoplanet sample, and 61 in our control sample), LOFTI was unable to generate a sufficient number of accepted samples (typically less than 5 accepted samples per 24 hours of runtime on a single CPU core). Often, when LOFTI is unable to find good orbital solutions like it, it is a sign that the system is not a real, physically bound binary, and may instead just be a chance alignment. Given the small number of these systems, the computational expense that would required to reach 10,000 orbits, and the fact that the systems are likely not physically bound anyway, we chose to exclude them from our final samples. 

\begin{figure}
    \includegraphics[width=\columnwidth]{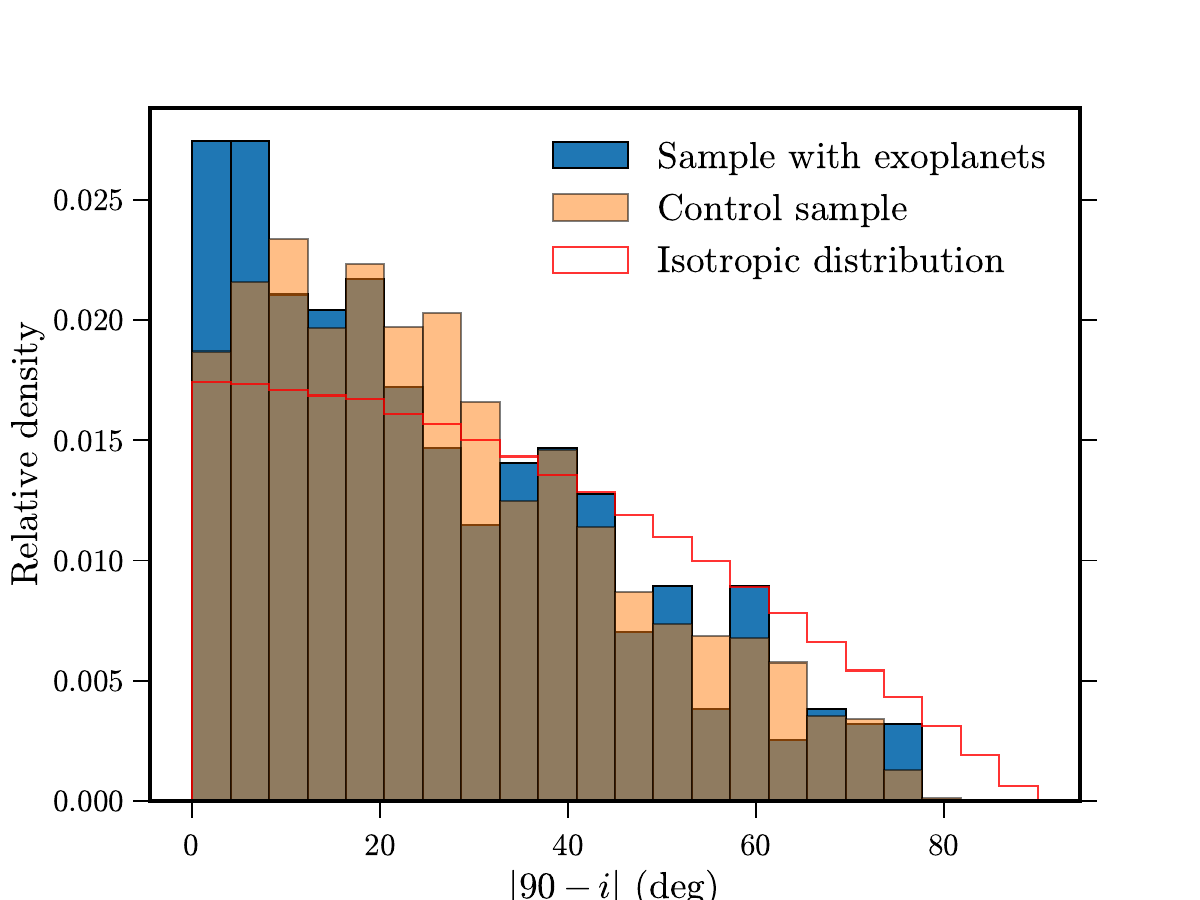}
    \caption{Histogram of the minimum inclination difference between binary orbits and the line of sight ($|90^\circ - i|$) for both a sample of binaries with exoplanets and a control sample without known exoplanets. An istropic distribution of inclinations generated using $p(i)\sim \cos(i)$ is also plotted in red. These histograms show our full sample, regardless of the planet radius and binary semimajor axis. Unlike \citet{Christian2022}, we see no clear evidence for preferential alignment in our full sample. We attribute this to the fact that the \citet{Christian2022} sample was primarily composed of small planets, which do show alignment (see Section \ref{sec:alignment}) while our full sample has many more hot Jupiters, which do not show a preferential alignment.}
    \label{fig:overallHist}
\end{figure}
\section{Results}\label{sec:results}
\begin{figure*}
    \includegraphics[width=\textwidth]{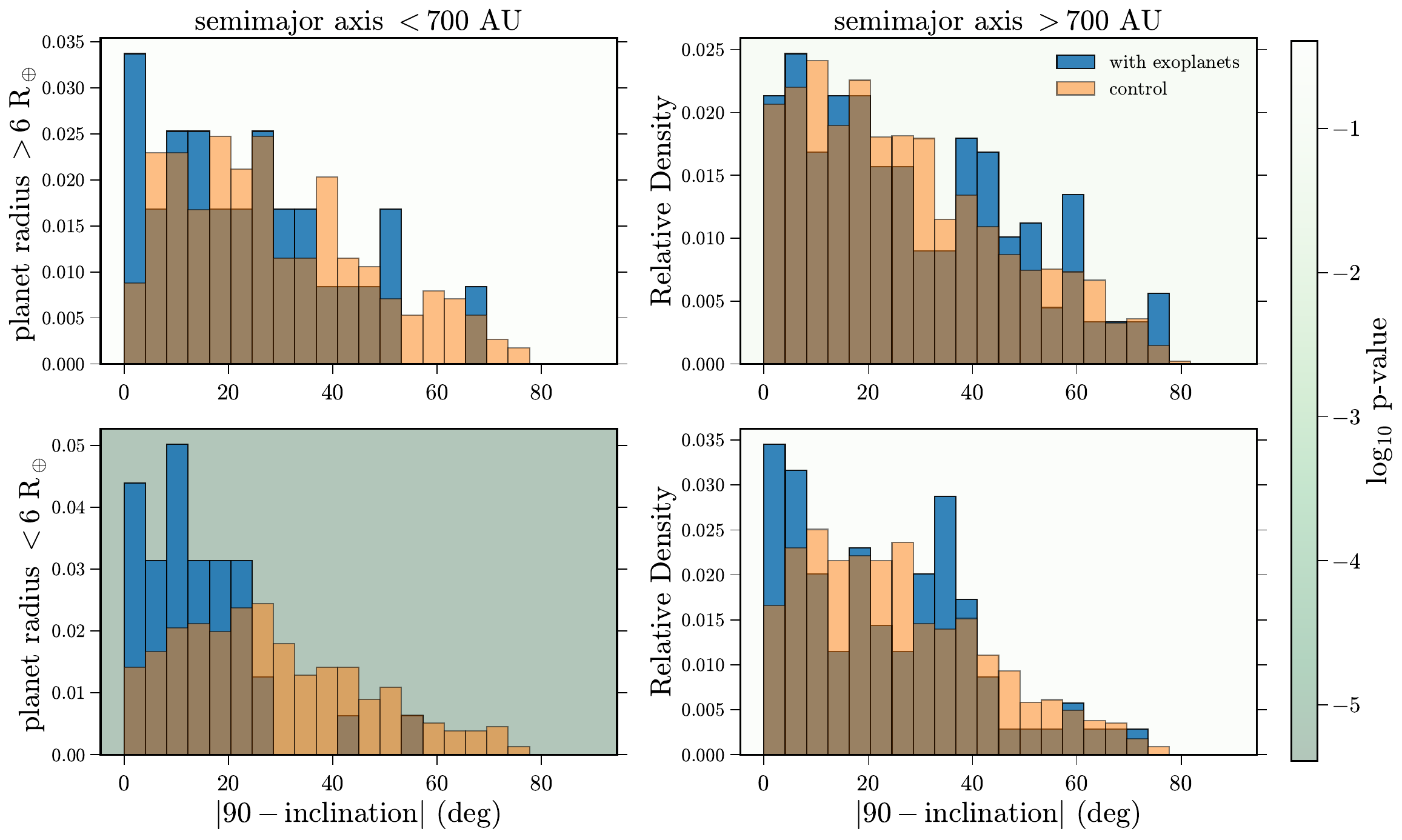}
    \caption{Four histograms of inclination binned by semimajor axis (left versus right) and radius (top versus bottom). The x-axis of each histogram is $|90-i|$. Orange represents the control sample, while blue represents the sample of binaries with exoplanets. The background color of each plot represents the p-value from the K-S test performed between control and the sample with exoplanets. We see strong statistical evidence for alignment in small planets in binary systems with small semimajor axes, but in none of the other regions. }
    \label{fig:fourplot}
\end{figure*}
\subsection{No Clear Preferential Alignment between Binary and Planet Orbits in the Full Sample}

Once we had completed our orbital modeling for each system in both the exoplanet sample and the control sample, we calculated $|90^\circ - i|$, the minimum misalignment between the orbit of a transiting exoplanet in each system and the binary orbit. In general, if there is alignment between planetary orbits and binary orbits, we would expect to find an overabundance of systems with $|90^\circ - i|$ near $0^\circ$ degrees for the exoplanet sample. We compared the distributions of $|90^\circ - i|$ for the exoplanet sample and the control sample, and show the resulting histograms in Figure \ref{fig:overallHist}. We also plot an isotropic distributions of inclinations.

While there is a hint of an overabundance of exoplanet-hosting systems near $|90^\circ - i| = 0^\circ$, the effect is considerably weaker than the preferential alignment seen by \citet{Christian2022} and is not statistically significant. A Kolmogorov-Smirnov test is unable to confidently reject the null hypothesis that the exoplanet sample's inclinations are drawn from a different population than the control sample's inclination  $(p=0.067)$. Both the control sample and the sample with exoplanets are significantly different from an isotropic distribution. This suggests that there are some underlying selection biases in our sample of binaries, so any statistical statement on inclination alignment should be made with respect to the control sample, not an isotropic distribution.

\subsection{Analysis of Sample Split in Planet Radius and Binary Semimajor Axis}

This lack of an overabundance of exoplanets with $|90^\circ - i| \approx 0^\circ$ is somewhat surprising given the evidence from multiple independent datasets \citep{Christian2022, Dupuy2022, Behmard2022, Lester2023AJ} of preferential alignment between planet and binary orbits in at least some parts of parameter space. We therefore investigated whether subsets of our sample do show a preferential alignment. In particular, \citet{Christian2022} noticed that the alignment they observed was strongest for closely orbiting binaries (with semimajor axes less than about 700 AU), while \citet{Behmard2022} found tentative evidence for misalignment of gas giant planets.  

We therefore chose to divide our sample along these two axes: planet radius (greater than or less than 6 \rearth) and binary semimajor axis (greater than or less than 700 AU). The divide of 6 $\rearth$ is chosen to roughly divide small and giant planets. We explore the sensitivity to the cutoff of semimajor axis cutoff in section \ref{sec:NoVisiblelargeplanets}.

The result of this is four subsamples, each containing between 30 and 86 objects. In a small number of cases, multi-planet systems host at least one planet smaller than 6 \rearth\, and at least one planet larger than 6 \rearth. In these cases, we chose to assign the system to both sub-samples. While this may raise concerns about double-counting, each of the sub-samples is analyzed independently, and we also confirmed that this strategy did not significantly affect any of the results in our paper. 

Figure \ref{fig:fourplot} shows histograms (similar to that shown in Figure \ref{fig:overallHist}) comparing the minimum misalignment between binary orbits and the line of sight for the control sample and for our exoplanet sample in each of the four sub-samples.

We calculate the Kolmogorov-Smirnov (KS) statistic for each of the four sub-samples and visually represent the p-value in the plots by shading the background color darker for lower p-values. No correction (e.g. Bonferroni correction or any of its variants) was applied for doing multiple comparisons. The derived p-values are $a<700$ AU, $r< 6 \rearth: 4 \times 10^{-6}$, $a<700$ AU, $r>6 \rearth: 0.4$, $a>700$ AU, $r< 6 \rearth$: $0.09$, $a>700$ AU, $r>6 \rearth$: 0.29.

The Anderson Darling test, a variant of the KS test which works better in some cases where the difference is at the edges of distributions, returns p-values of $a<700$ AU, $r< 6 \rearth: 0.001$, $a<700$ AU, $r>6 \rearth: 0.2$, $a>700$ AU, $r< 6 \rearth$: $0.1$, $a>700$ AU, $r>6 \rearth$: 0.2. For the remainder of this paper, we will use the KS statistical test, since the results are qualitatively similar when using the Anderson Darling test.



\begin{figure*}
    \includegraphics[width=0.9\textwidth]{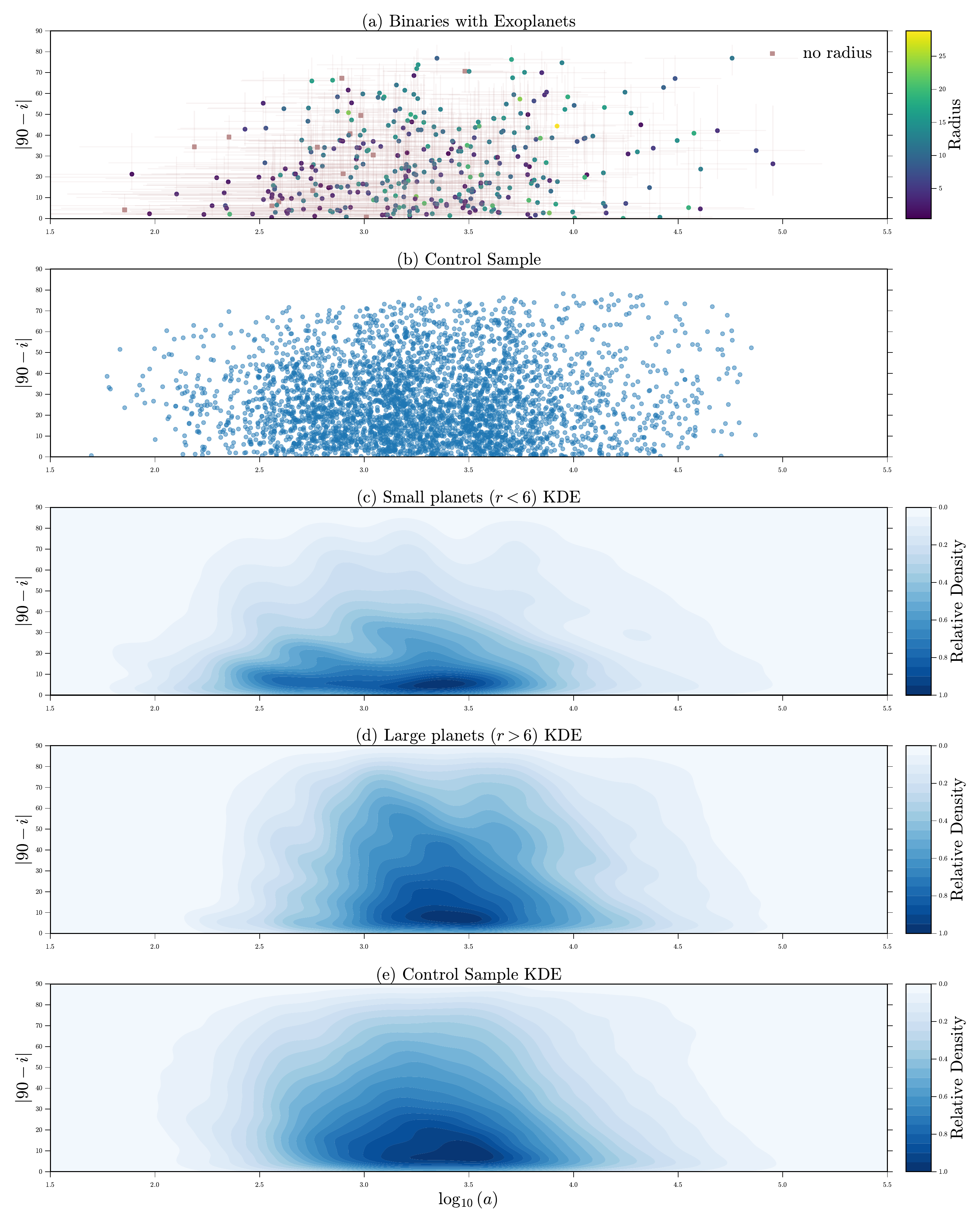}
    \caption{(a) Plot of median inclination vs. median semimajor axis for binaries with exoplanets from \Gaia\ DR3. Some exoplanets do not have reported radii values so are represented as maroon squares. Since the errorbars on inclination are the 16th and 84th percentiles of $|90-i|$, while the median values are reported as $|90-\overline{i}|$ where $\overline{i}$ is the median value of inclination, occasionally an errorbar will go under zero degrees in which case we cut it off. (b) The same as (a), but for the control sample. (c), (d) are KDEs of systems with planets that have with radii greater than or less than $6 \rearth$, respectively. (e) is a KDE of the control sample systems in panel (b). The KDEs are drawn from 100 random samples from each binary to reflect the underlying error present in the sampling procedure. The precise difference in the distributions of various distributions is best seen in Figure \ref{fig:moreComp}.}
    \label{fig:incSemi}
\end{figure*}

\subsubsection{Strong Alignment in Small Planet Systems with Binary Semimajor Axes Below 700 AU} \label{sec:alignment}


Figure \ref{fig:fourplot} shows a strong alignment between the orbits of small planets and binary companions with semimajor axes less than about 700 AU (lower left panel). 
The Kolmogorov-Smirnov test rejects the null hypothesis that the inclinations of the control sample binaries are drawn from the same population as the inclinations of close binary systems containing small planets (with p-value of $1.2 \times 10^{-6}$). This finding confirms the previous hints from \citet{Christian2022} and \citet{Behmard2022}, and increases the level of statistical significance by several orders of magnitude.







\begin{figure*}
    \includegraphics[width=0.9\textwidth]{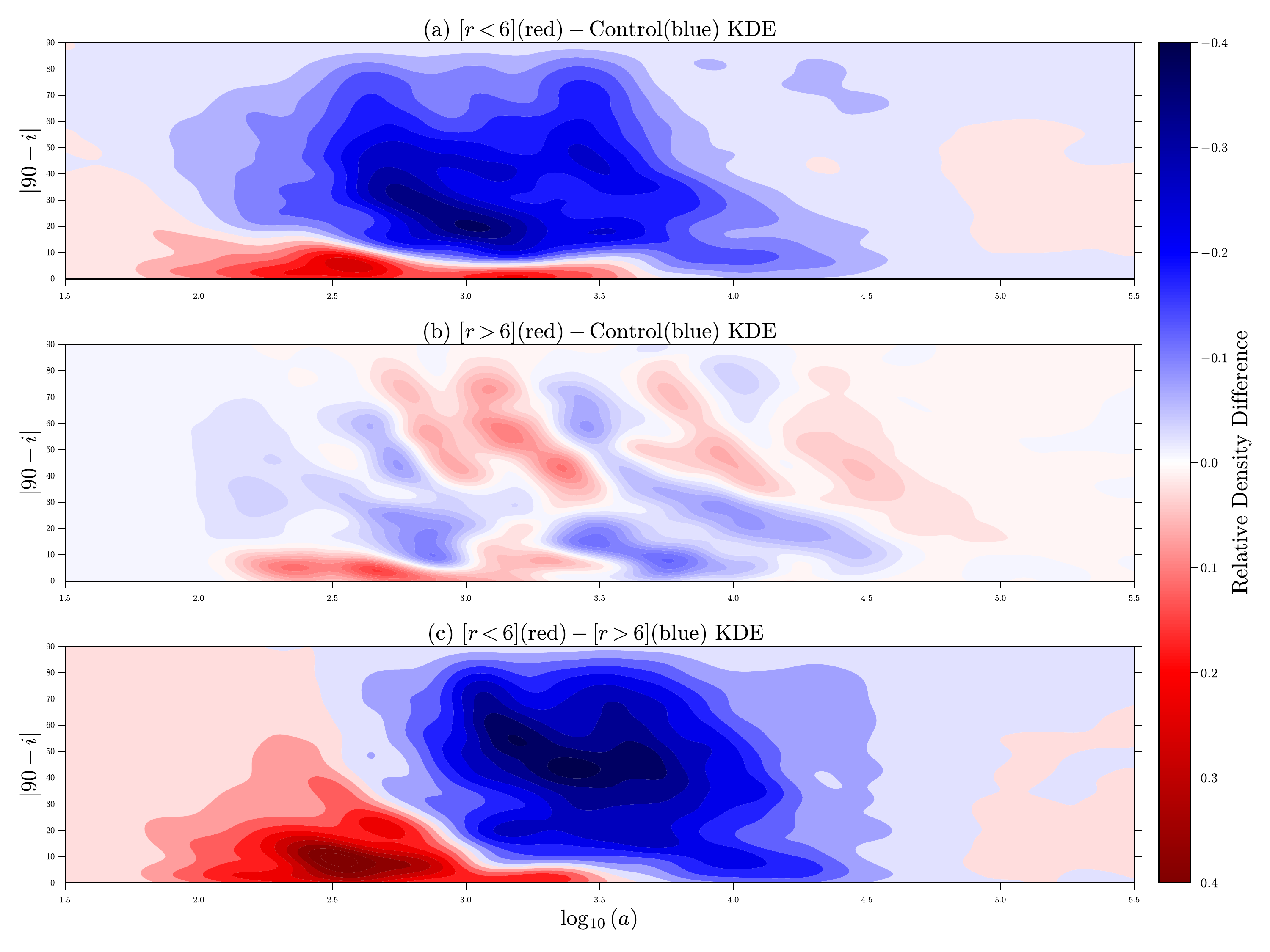}
    \caption{A similar arrangement to Figure \ref{fig:incSemi} (c-e) except comparing differences between populations. The relative density difference is calculated as the difference in the normalized density of Figure \ref{fig:incSemi} between two different populations. (a) shows binaries with $r<6 \rearth$ compared to the control sample, (b) the same but with $r>6 \rearth$, and (c) comparing $r<6 \rearth$ and $r>6 \rearth$. For example, darker red in (a) indicates relatively more small planets compared to the control sample, while lighter colors mean that the small planet and control sample have similar densities. The relative biases in the semimajor axis distributions of large and small planets are apparent in panel (c). Because larger planets are found at further distances, the semimajor axes of the binary systems they are found in is systematically larger.}
    \label{fig:moreComp}
\end{figure*}

The analysis shown in Figure \ref{fig:fourplot} shows with high confidence that there is preferential alignment between the orbits of small planets with semimajor axes less than 700 AU and the orbits of the binary systems they reside in, but this does not take into account the uncertainties in both the binary inclination angle and semimajor axis, which in many cases are fairly large. We therefore performed analysis to visualize the impact of uncertainties on these parameters. We randomly sampled 100 points from the posterior probability distributions for the binary inclination angles and semimajor axes for each system in both the exoplanet sample and the control sample. We calculated a 2D Kernel Density Estimate (KDE) of the mutual distribution of these parameters for both samples, which we show in Figure \ref{fig:incSemi}. Subtracting the KDEs of the two samples (Figure \ref{fig:incSemi}, panel e) reveals a significant excess in probability mass at near edge-on binary orbits and semimajor axis less than 1000 AU. Evidently, the pattern we noticed based on the distribution of median values of a given parameter for each system still exists when taking uncertainties on these parameters into account.  We also see evidence of a dichotomy between small planets and large planets when viewing the sample in this manner (see Section \ref{sec:NoVisiblelargeplanets} for more details). Figure \ref{fig:moreComp} shows the same KDEs, but now split between small planets (see Figure \ref{fig:moreComp} panel a, which show a strong excess in probability mass for edge-on orbits at small semimajor axes) and large planets (see Figure  \ref{fig:moreComp}, panel b, which shows no clear excess at edge-on orbits, but some preference for misaligned orbits with respect to the control samples at semimajor axes between 1000 and 10,000 AU). Finally,  Figure  \ref{fig:moreComp}, panel c shows the difference between panels a and b, highlighting the overabundance of well-aligned systems with small planets.


\begin{figure*}
    \includegraphics[width=\textwidth]{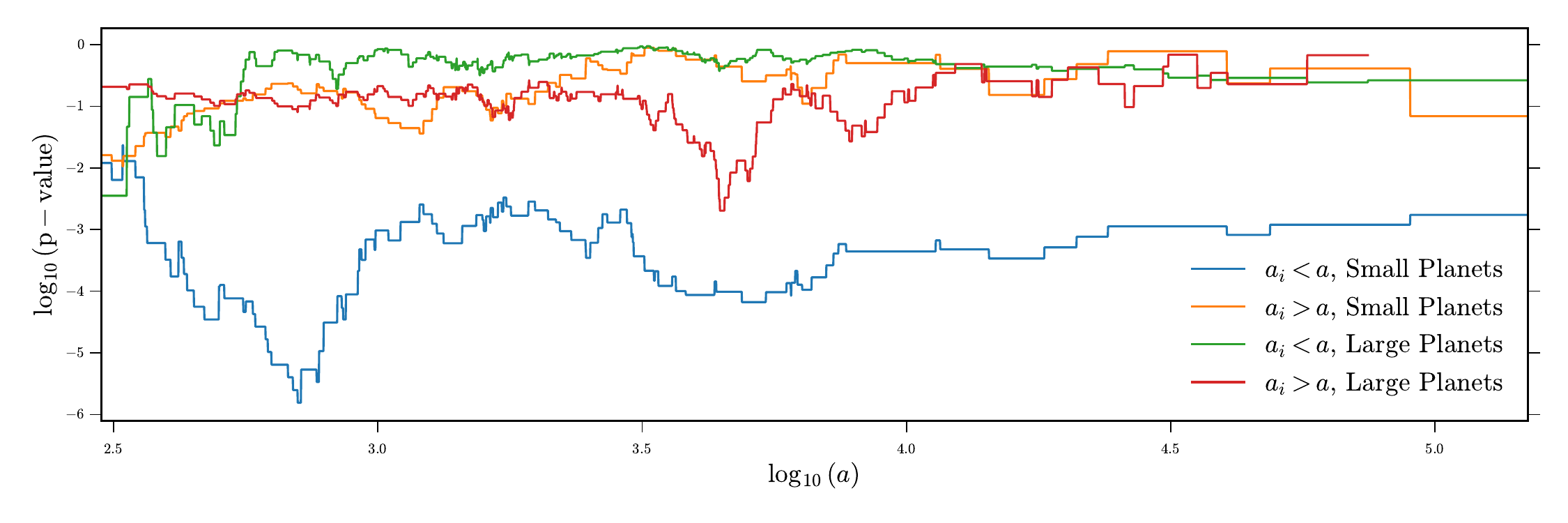}
    \caption{The variation of the p-value between binaries with exoplanets above and below a specific semimajor axis and the control sample, as measured through a KS test. Explicitly, denoting the control sample as $c$ and the binaries with exoplanets as $b$, the blue line is computed as $\text{KS}(b[a_i<a\text{ and } r<6],c)$, the orange line as $\text{KS}(b[a_i>a\text{ and } r<6],c)$, the red line as $\text{KS}(b[a_i<a\text{ and } r>6],c)$, and the green line as $\text{KS}(b[a_i>a\text{ and } r>6],c)$. Because we are doing many comparisons, per \citet{Schlaufman2015ApJL}, the minimum of the p-value distribution should not be interpreted as an exact probability, but the dip at $\approx$ 700 AU suggests that there is some change in the inclination distribution around roughly that semimajor axis.}
    \label{fig:pvalueSemi}
\end{figure*}

\subsubsection{No Clear Evidence for Alignment in the Largest-Separation Binary Systems}
\label{sec:NoVisible}

\citet{Christian2022} found that the alignment they observed in their sample was largely driven by systems with binary semimajor axes less than 700 AU. We find a similar result. Neither of the sub-samples with binary semimajor axes greater than 700 AU show statistically significant evidence for alignment. The Kolmogorov-Smirnov test yields $p = 0.29$ for the small planets and $p = 0.09$ for the large planets. \citet{Christian2022} interpreted the fact that alignment is strongest at semimajor axes less than 700 AU as tentative evidence that dynamical realignment of protoplanetary disks by the binary companions may explain the pattern they observed. We reassess this hypothesis in light of our increased sample size in Section \ref{sec:dynamics}.  

We note that our nominal value of 700 AU for the cutoff binary semimajor axis below which alignment becomes strong is not a precisely measured value. \citet{Christian2022} identified this cutoff by visually inspecting a plot of minimum alignment versus binary semimajor axis, and we use this value to split our sample for the sake of consistency with their analysis. We wish to emphasize that our choice to split the samples at 700 AU should not be taken as evidence that there is a physical cutoff at precisely 700 AU, or that indeed there are two distinct populations of systems separated by some semimajor axis at all. Despite the increased sample size over our previous paper, there are still only 40 small planets with semimajor axes below 700 AU, so determining the exact location of the cutoff is difficult. 

We performed an analysis inspired by \citet{Buchhave2014Natur} to assess the most likely location of a transition at which binaries are preferentially aligned with short-period planets. We took all four of our subsamples and repeated our analysis calculating the p-value of a Kolmogorov-Smirnov test between our exoplanet and control sample when we split the samples at different binary semimajor axis values. The p-values we measured are plotted against the semimajor axis above which we excluded systems from our calculation in Figure \ref{fig:pvalueSemi}. We indeed see a minimum in the p-value for small planets with small binary separations around 700 AU, but the minimum is broad; from this, we infer that the cutoff semimajor axis above which alignment disappears, if it exists, is unlikely to be above $10^3$ AU. We note that the decrease in p-value for small planets past $10^{2.5}$ AU is mainly due to an increasing sample size rather than different sample characteristics at those semimajor axis values. Per \citet{Schlaufman2015ApJL}, we caution against reading too much into the exact minimum p-value in Figure \ref{fig:pvalueSemi}, and we note that it is also entirely possible that there is instead a gradient of systems from aligned to not aligned that exists somewhere around 700 AU. 

There is also a dip in the p-value when separating large planets above and below 5000 AU. The p-value at this dip is less than 0.01, but given the many comparisons taking place across the range of semimajor axis cutoffs, this cannot be interpreted to be statistically significant.



\subsection{No Clear Evidence for Anti-Alignment in Large Planet Systems}
We inspected the histograms for large planets (with radii greater than 6 \rearth). Visually, there is an apparent overdensity in the number of systems with semimajor axes greater than 700 AU with minimum inclination differences of about 35 to 45 degrees. This feature is intriguing at first glance because this is close to the maximum of the inclination range of Lidov-Kozai oscillations (see e.g. \citealt{Naoz2016}), but there is no evidence this is a statistically significant feature. Indeed, when we instead plot the histogram of $\sin(|90-i|)$, where the isotropic distribution takes on the form of a uniform distribution, the excess does not not stand out above the noise in the histogram, and no longer appears visually statistically significant. 


\subsubsection{No Clear Evidence for Alignment in Large Planet Systems}
\label{sec:NoVisiblelargeplanets}

Finally, we see no clear evidence for alignment between wide binaries with semimajor axes below 700 AU, and large planets in Figure \ref{fig:fourplot}. The Kolmogorov-Smirnov test yields a p-value of 0.4 for this subsample. This is in contrast to the strong alignment we see between the orbits of small planets and wide binaries with semimajor axes less than 700 AU.  The lack of a statistically significant alignment for large planets at close semimajor axes could be the result of either an astrophysical difference between small and large planet systems, or it could be an artifact of the samples involved (either the large planet sample is too small for adequate comparisons, or it is contaminated). We therefore strove to determine which is the case. 

First, we considered whether the samples are simply too small to make definitive comparisons. The sample of large planets at small semimajor axes contains only $\approx 30$ systems, and the sample of small planets at small semimajor axes is not much larger ($\approx 40$). This makes it difficult to directly compare the two distributions with high statistical confidence. We tested whether the sample size is sufficient to confidently distinguish the two inclination distributions by performing a KS test directly comparing the large and small planet samples. On face value, the difference in distributions of close-in large and small planets is border-line statistically significant, with a KS test p-value of 0.012, implying that the two distributions do seem significantly different, and that the samples are indeed large enough to draw conclusions about differing degrees of alignment. 

Second, we considered the more vexing possibility that the difference between the two samples is caused by a different false positive rate between the two samples, and therefore different levels of false positive contamination. The vast majority of the large planets in our sample are hot or warm Jupiters, which are intrinsically rare, and are difficult/impossible to distinguish from other astrophysical scenarios (like eclipsing brown dwarfs and low-mass stars) from TESS data alone. Therefore, the TESS false positive rate for these planet candidates ($\approx$ 35\% for hot Jupiters in particular) is considerably higher than for small planet candidates \citep[$\approx$ 15\%;][]{Zhou2019}. One mitigating factor for our study is that these rates are probably overestimates, since we have already discarded a number of false positives from our sample thanks to reconnaissance ground-based observations. 72 out of 250 of the hot Jupiters in our sample are confirmed planets, and more are suggestively planets from follow-up observations. Since many planets are already confirmed (or confirmed as false-positives), a 35\% false-positive rate is a high upper bound for the true false positive rate.

In an attempt to quantify the effects of the larger false positive rate of large planets, we devise a bootstrap test based on the sample of small planets. Our analysis seeks to answer the question: \textit{``If the inclination distribution for large planets were drawn from the same inclination distribution for small planets, would false positive contamination cause us to see as large of a difference between the populations as we observe?''} To do this, we performed a test where we drew sample inclinations from the distribution of inclinations we found for small planets, added in some fraction of randomly chosen inclinations from our control sample, and calculated the KS statistic between the new, simulated sample of giant planets (with false positives) and the control sample. In recognition of the strong dependence on the assumed false positive rate and the inexact nature of the cutoff at about 700 AU in binary separation, we perform these simulations for a variety of different values of these parameters. 

In particular, we took the following steps: 
\begin{enumerate}
\item We selected the false positive rate ($X$) and the semimajor axis cutoff ($Y$) for a particular bootstrap test. The procedure is computationally taxing, so we computed the bootstrap procedure on a grid of only 20 false positive points and 500 semimajor axis points.
\item Given the chosen semimajor axis cutoff $Y$, we counted the number of systems with planets larger than 6 $R_\oplus$ and semimajor axis less than $Y$. We call this number of large-planet systems $A$. 
\item We then identify the sample of small planet systems with semimajor axes less than $Y$. 
\item We randomly select $\lfloor A(1-X)\rfloor$ systems from the sample of small planet identified in the previous step and record the inclinations of these systems. Here, $\lfloor \rfloor$ denotes the ``floor'' operator since only an integer number of samples can be used in a given comparison. These predominantly well-aligned binary inclinations simulate a well-aligned distribution of large planet systems. 
\item We then randomly select $\lceil AX\rceil$ systems from the control sample and record the inclinations of these systems, adding them to the sample of inclinations identified in the previous step, and bringing the total number of inclinations in the synthetic sample to $A$. These randomly aligned systems from the control sample simulate the effects of a population of false positives with no preferential alignment among the hot Jupiters. Here,  $\lceil \rceil$ denotes the ``ceiling'' operator. 
\item We calculate the KS statistic comparing the simulated population of giant planets (with false positives mixed in) and the control sample for giant planets, and record the p-value. We repeat this test 10,000 times for each value of $X$ and $Y$, and record the median p-value and fraction of the tests where the p-value is below 0.05 for each $(X,Y)$ pair. 
\end{enumerate}

\begin{figure*}
    \includegraphics[width=\textwidth]{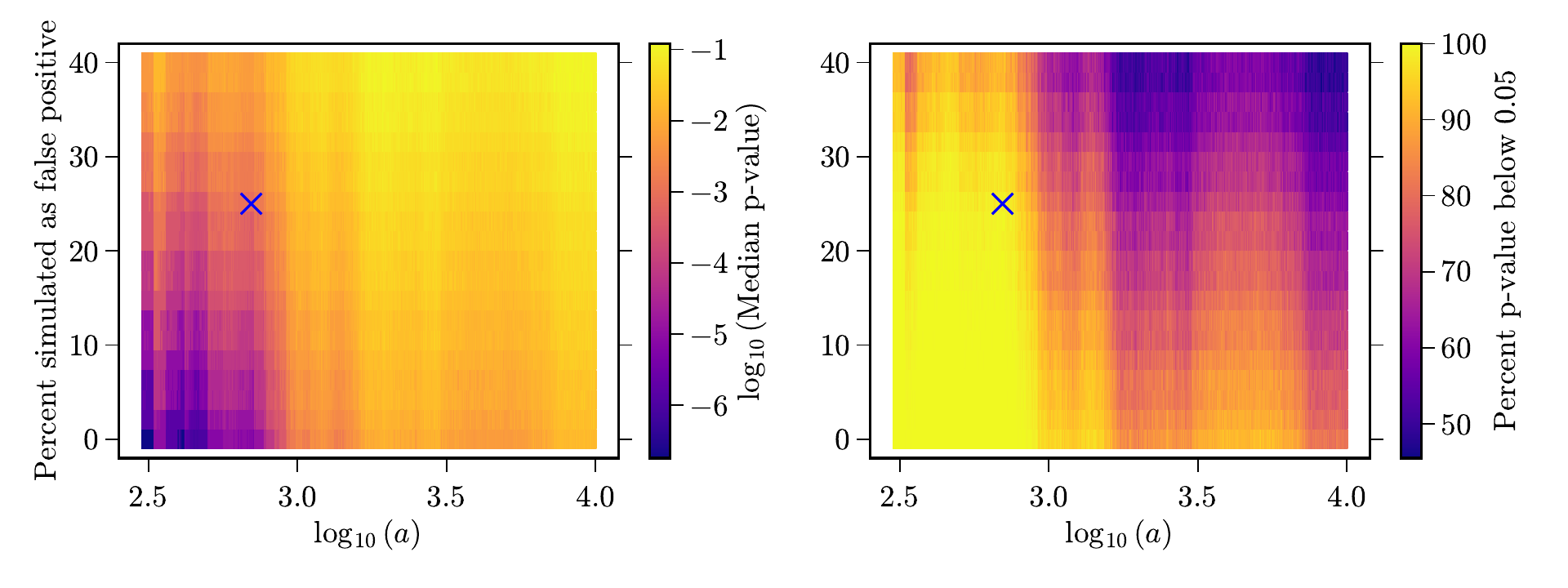}
    \caption{Results of simulations to determine whether higher false positive rates for giant planets can explain the difference in the inclination distribution compared to small planet systems (described in Section \ref{sec:NoVisiblelargeplanets}). \textit{Left:} an image showing the median p-value from 1000 simulations performed for a given false positive fraction and semimajor axis cutoff. \textit{Right:} an image showing the percentage of p-values from each of the 1000 simulations that are below 0.05. The horizontal and vertical lines visible in the figure are the separations between the grid points used in the simulation. In both plots, a blue x marks our expectation for the semimajor axis delineation and false-positive rate. Our grid points have a much finer resolution in semimajor axis than false positive rate. We find that for plausible values of the false positive rate (expected to be roughly less than 25\%) and semimajor axis cutoff ($\approx$ 700 AU), we would probably still expect to see a significant alignment if giant planets were similar to small planets. Since we do not see a significant alignment, we infer that giant planets are probably less strongly aligned than small planets.}
    \label{fig:bootstrap}
\end{figure*}

Figure \ref{fig:bootstrap} shows the results of these simulations for a variety of different false positive rates and semimajor axis cutoffs. The left-hand panel shows the median p-value for the suite of tests, and the right-hand panel shows the fraction of the tests where the p-value is below 0.05. As expected, for low false positive rates (where the fewest misaligned false positive systems are mixed in) and small values of semimajor axis cutoff (where the alignment is strongest), we see low median p-values and high fractions of the time yielding a significant difference between the simulated sample and the control sample. In particular, for the nominal values for these parameters (roughly 25\% false positive fraction and 700 AU cutoff), we expect that we should usually see a statistically significant difference between the exoplanet sample and the control sample. For injections with a cutoff of 700 AU and false positive rates less than or equal to 25\%, the minimum of percentage of samples in a given injection with p-value less than 0.05 is 90\%, a p-value of 0.1 is 96\%, a p-value of 0.2 is 98\%, and a p-value of 0.3 is 99\%. In reality, we see a KS test p-value when we compare the exoplanet and control sample considerably higher than these typical values (0.4), implying that false positive contamination most likely cannot explain the difference between the small planets and large planets. This suggests that large planets in binary systems are probably not aligned to the same degree as small planets.

A 25\% false-positive rate does not include the built-in false-positive rate for small planets. If, say, the small planet rate is 15\%, then this would correspond to a 40\% false-positive rate for hot Jupiters. As a reminder, follow-up observations have already confirmed that some of the hot Jupiters in our sample are planets, so a simulated false-positive rate of 25\% is a generous upper bound.

\subsection{No dependence on temperature of the host star}

\begin{figure*}
    \includegraphics[width=0.9\textwidth]{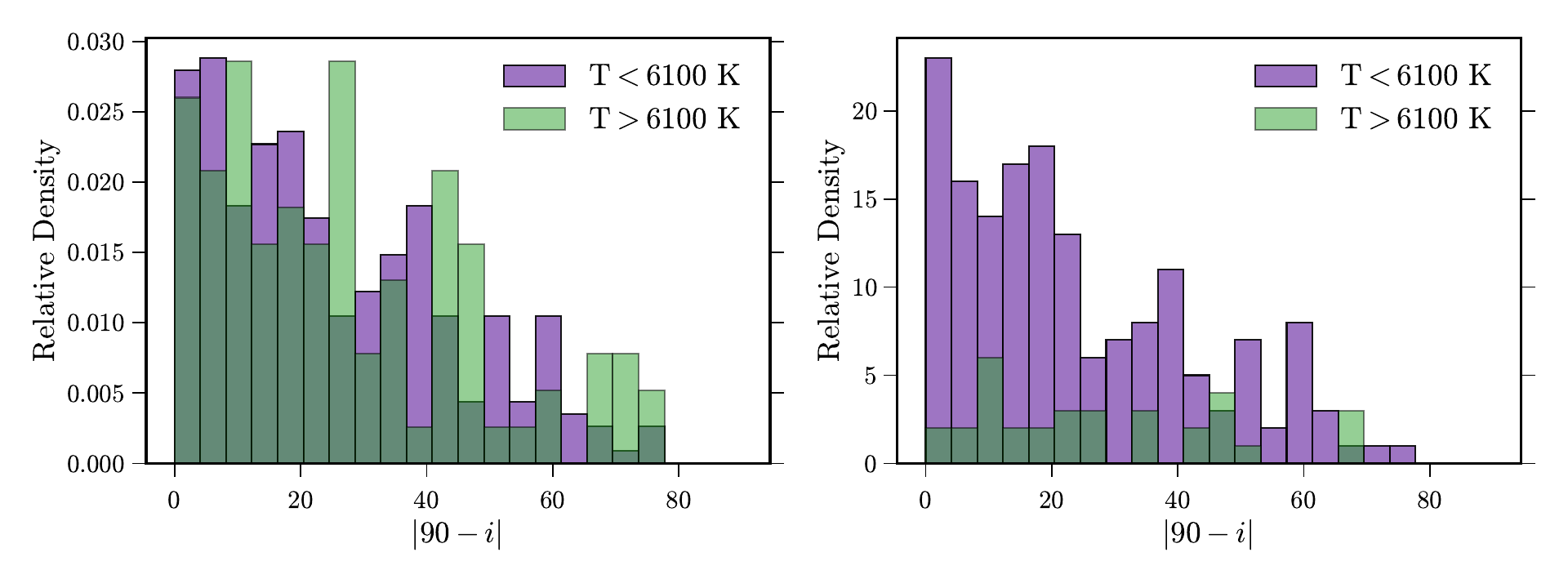}
    \caption{\textit{Left:} The distribution of $|90-i|$ for all systems with exoplanets when the temperature of the host star of the exoplanet has a temperature greater than or less than 6100 K, respectively. Unlike the distribution of spin/orbit angles for hot Jupiters, we find no evidence of a difference in the distribution of planet/binary orbit angles above and below the Kraft Break. \textit{Right:} The same as the left but restricted to less than 700 AU, and with counts shown instead of relative density.}
    \label{fig:tempDist}
\end{figure*}

Finally, we tested whether the degree of alignment between planet and binary orbits depends on the temperature of the host star in question. We are motivated to perform this test because temperature of the host star is an important predictor of whether the spin/orbit angle (obliquity) of hot Jupiter systems is aligned or misaligned. Famously, hot stars above the Kraft break ($\sim 6100$ K) host hot Jupiters that span a wider range of obliquities than cool stars \cite[e.g.][]{winn2010, Schlaufman2010,Albrecht2012ApJ}. More recently, \citet{rice2024} suggested that the orbits of close-in binary star and exoplanet orbits should be fully aligned at small semimajor axes for temperatures below the Kraft break, and less alignment should be seen above the Kraft break.  However, we find no dependence of the alignment between planet orbits and wide binary orbits on the host star's temperature, either for our full sample, or only systems with large planets (dominated by hot Jupiters). We plot the distribution of inclinations for host stars with temperatures less than and greater than 6100 K in Figure \ref{fig:tempDist}.  For all stars, a KS test between host stars with temperatures less than and greater than 6100 K returns a p-value of 0.55. For stars with large planets, the KS test gives a p-value of 0.80 and for small planets, a p-value of 0.54. We attribute the lack of a statistically significant difference here to the very small number of hot stars with close-in binary companions ($a<700$ AU) in our sample. There are only 9 systems in our sample with $T>6100$ K and $a<700$ AU.


\section{Discussion} \label{sec:discussion}

\subsection{Comparison to Previous Work}

The results of this work are largely in agreement with those of previous papers \citep{Christian2022, Dupuy2022, Behmard2022, Lester2023AJ}, but strengthen and add some nuance to their conclusions. 

\begin{figure*}
    \includegraphics[width=.9\textwidth]{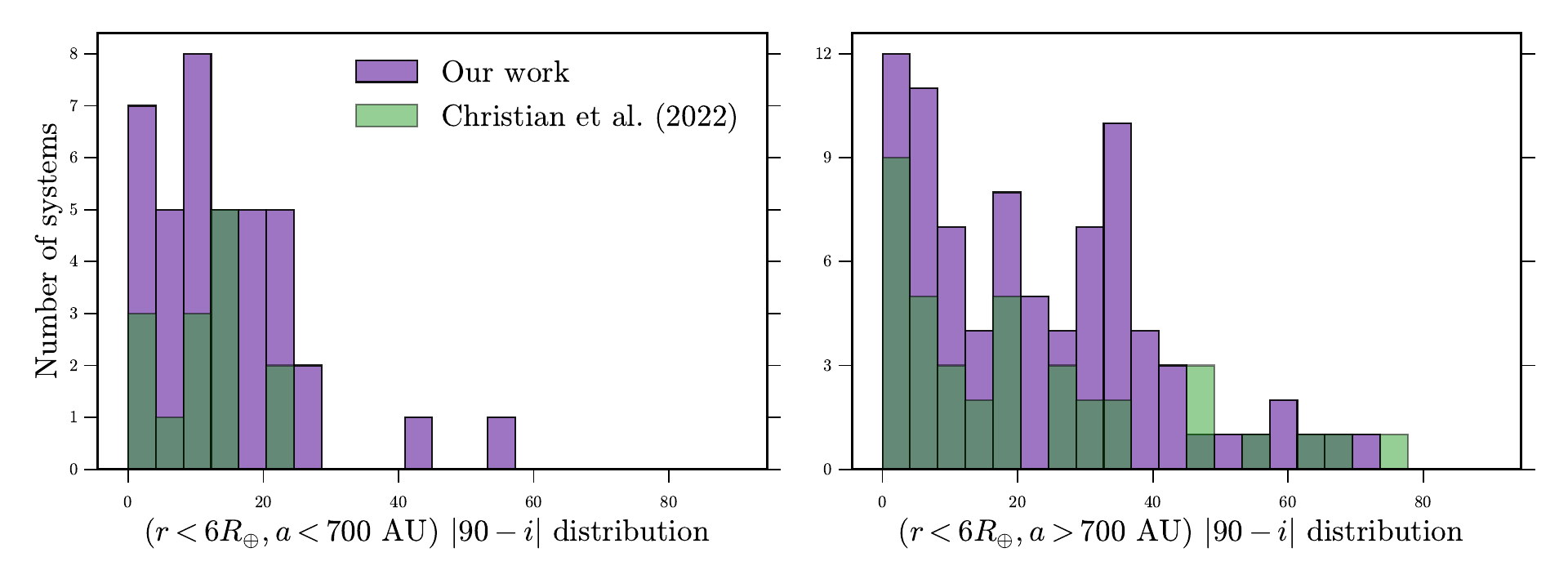}
    \caption{\textit{Left: } Comparison of the distributions of binary inclination for systems with small planets with binary semimajor axes less than 700 AU (\textit{Left}) and with binary semimajor axes greater than 700 AU (\textit{Right}) between our sample and that of  \citet{Christian2022}.  In both panels, our results are consistent with those of \citet{Christian2022}, but our considerably larger sample lets us make inferences with higher statistical confidence.
    }
    \label{fig:previousWork2}
\end{figure*}

Previously, \citet{Christian2022}, found that wide binaries (binary systems with semimajor axes greater than 100 AU) and planets of all sizes showed statistically significant alignment, and that the alignment was most prevalent in systems with binary semimajor axes less than 700 AU. Now, thanks to a considerably larger sample, with a greater diversity of planetary types, we see the alignment identified by these previous works is predominantly in systems containing small planets and binary systems with small semimajor axes. This is in good agreement with the results of both \citet{Christian2022} and \citet{Dupuy2022}, who studied closer binary systems than \citet{Christian2022} and found a higher degree of alignment than in the overall \citet{Christian2022} sample. In Figure \ref{fig:previousWork2}, we plot distributions of inclination for our sample compared to \citet{Christian2022}. We quantified the consistency between our sample and that of \citet{Christian2022} by performing a KS tests comparing the two samples. A KS test comparing the distribution of small planets with $a<700$ AU in our sample versus that of \citet{Christian2022} returns a p-value of 0.50, and a KS test comparing the samples for small planets with $a>700$ AU yields a p-value of 0.59, indicating good consistency.

We also are able to corroborate the tentative evidence found by \citet{Behmard2022} that hot Jupiter systems do not show the same alignment as small planet systems, and in fact may be misaligned. With a larger sample than \citet{Behmard2022}, we also see no evidence for alignment for systems with large planets, and see tentative evidence for preferential misalignment in Figure \ref{fig:moreComp}. These findings should continue to strengthen  as the planets discovered by TESS in this sample continue to receive follow-up observations and mitigate the risk that false positive contamination is affecting our results. 



\subsection{Implications for dynamical conclusions} \label{sec:dynamics}

\citet{Christian2022} suggested that a disk torquing mechanism, influenced by energy dissipation in the protoplanetary disk due to a binary companion, would result in alignment of the orbits of planets forming the disk with the binary orbit. This seemed to explain the observed alignment cutoff for planets of all masses at approximately 700 AU.

However, our new expanded data set reveals a contrast not apparent in our previous smaller sample: while systems containing smaller planets exhibit alignment within this range, systems containing large planets (predominantly hot Jupiters\footnote{We note that while our large-planet sample nominally contains both hot Jupiters and warm Jupiters, we treat it as dominated by hot Jupiters. We attempted to test whether hot Jupiters and warm Jupiters showed different behavior by separating them using the definition of \citet{huang2016}, but there too few (only 23) warm Jupiter systems to draw definitive conclusions. Adding warm Jupiter systems from \Kepler\ and from ongoing TESS discoveries as the survey baseline increases may enable such analysis in the future.}) do not. This suggests that systems containing small planets and those containing hot Jupiters undergo different evolutionary paths. Two potential explanations emerge: dynamical excitation of star-planet misalignments during the late phases of the disk lifetime and disk dissipation, or subsequent dynamical evolution of the hot Jupiter orbit over much longer timescales after the disk has dissipated. 

First, the picture described in \citet{Christian2022} of disk-binary orbits being aligned due to energy dissipation does not account for the potentially destabilizing effect of secular resonances, which may occur when the fundamental frequencies mediating the interactions in the planetary system evolve due to dissipating disks \citep{Spalding2014, EpsteinMartin2022}. 
When two precession frequencies in the system become commensurate due to this evolution, it can lead to large-scale excitation of orbital parameters. 
In particular, systems containing hot Jupiters, where a large amount of angular momentum is concentrated at one orbital radius, may be strongly affected by secular resonances as their parent disks dissipate \citep{Zanazzi2023}. This mechanism can operate even if hot Jupiters form \emph{in situ} \citep{Batygin2016}. Such resonant excitations occur more readily in systems hosting Jupiter-mass planets, and would not operate as efficiently to create misalignment in systems with smaller planets.


Second, hot Jupiters have multiple possible formation models. While some hot Jupiters are thought to form either \emph{in situ} or via dynamically quiet disk migration \citep[e.g.][]{Becker2015ApJL, Zhu2018RNAAS,  Hord2022AJ, Sha2023MNRAS}, recent evidence suggests that the majority of hot Jupiters form via high-eccentricity migration \citep{Zink2023}, a process which requires an initial dynamical excitation due to a companion and subsequent orbital circularization due to tides raised on the planet's and star's surfaces \citep{Dawson2018}. 
Many models of hot Jupiter migration invoke an inclined companion to start the initial migration process, and many of these mechanisms start with an inclined companion and subsequent Lidov–Kozai oscillations which perturb the Jupiter into a small-periastron orbit that can then tidally circularize \citep{Wu2003, naoz2012, Petrovich2015, Naoz2016, Vick2019}.  This picture would produce the lack of binary companion alignment seen in the systems hosting hot Jupiters in our sample, as small planets do not have similar origin pathways. 
However, Lidov-Kozai oscillations likely explain the origin of only a small fraction \citep[20\% or less;][]{Petrovich2015} of the known hot Jupiters. Instead, \citet{Zink2023} suggests that the coplanar high-eccentricity migration mechanism of \citet{Petrovich2015b} is likely dominant, meaning that even initially coplanar outer companions can perturb cold Jupiters into hot orbits, where they are more readily discovered.

\citet{winn2010} and \citet{rice2022} find that there is a population of cool stars hosting hot Jupiters with low stellar obliquities. This population generally comprises most observed hot Jupiters \citep{Dong2023}, and likely composes a significant fraction of our systems given that 72\% of the hot Jupiters in our sample orbit cool stars. The hot Jupiters with cool stars probably have predominately aligned stellar spins and planetary orbits, but we observe a distribution of inclinations consistent with isotropy. These two observations imply that the distributions of stellar spins and binary orbits should be roughly uncorrelated.

However, one puzzle remains: under the assumption that we are mostly observing this population, if we were to measure the obliquity (stellar spin/planet orbit) angle of the systems in our sample, it should be uncorrelated with the inclination angle (planet orbit/binary orbit). This is in tension with \citet{Behmard2022}, who found tentative evidence for correlations between obliquity and inclination distributions. However, \citet{rice2024} used a larger sample of systems and did not find evidence for correlations between obliquity and inclination distributions. 

Spin-orbit measurements of the stars hosting hot Jupiters in our sample could help resolve this uncertainty and determine the angular momentum hierarchies in these hot-Jupiter hosting systems, enabling a more concrete conclusion on the dynamical origins of these hot Jupiters.




\section{Conclusions}\label{sec:conclusions}
Previously, several studies \citep{Christian2022, Dupuy2022, Behmard2022, Lester2023AJ} found evidence that the orbits of exoplanets tend to be aligned with the orbits of wide binary companions in these systems. We extend the work of \citet{Christian2022} in particular, significantly increase the size of the sample, and update their conclusions. We do this by determining the probability distributions of the orbital parameters of binary systems that host transiting exoplanets (from the TESS mission) using astrometric measurements from \textit{Gaia} DR3.

 We confirm and amend the findings of these previous papers and show that the alignment seen previously is dominated by systems containing small planets (smaller than about 6 $R_\oplus$) in binary systems with semimajor axes less than about 700 AU. The significance of the alignment we detect is considerably greater than that reported by previous studies. 

One interesting result of our work is that at first glance, it appears that our sample of large planet systems (dominated by hot Jupiters) in wide binaries are not preferentially aligned to the same degree as small planets. One must be careful in confidently claiming this to be true given several selection biases, most importantly the relatively high false positive rate of hot Jupiters (Section \ref{sec:NoVisible}). Despite these hurdles, a bootstrap analysis shows that the higher false positive rate of hot Jupiters compared to small planets most likely cannot explain the difference in alignment between the two samples. This result suggests that the orbits of hot Jupiters in relatively close ($a\lesssim700$AU) visual binary systems show less alignment with the binaries than similar systems containing small planets (Section \ref{sec:NoVisiblelargeplanets}). This is likely the signature of the dynamical upheaval believed to be associated with hot Jupiter migration.

As time goes on, it should be possible to continue strengthening the conclusions from this work. Future \Gaia\ data releases should result in more discoveries of binary systems and better constraints on their orbital parameters. Meanwhile, as TESS continues to observe, and astronomers follow-up the planet candidates and eliminate more false positives, the purity and size of the TESS planet sample should improve. We expect that as our sample grows and improves, the question over the difference in small planet and hot Jupiter alignment will be resolved with greater certainty.

\section*{Acknowledgements}

We acknowledge helpful conversations with Bekki Dawson. S.C. acknowledges support from the MIT Undergraduate Research Opportunities Program (UROP).  The authors acknowledge the MIT SuperCloud and Lincoln Laboratory Supercomputing Center for providing high-performance computing resources that have contributed to the research results reported within this paper. F.J.P acknowledges financial support from the grant CEX2021-001131-S funded by MCIN/AEI/ 10.13039/501100011033 and through projects PID2019-109522GB-C52 and PID2022-137241NB-C43.
This paper includes data collected by the TESS mission, which are publicly available from the Mikulski Archive for Space Telescopes (MAST). Funding for the TESS mission is provided by NASA’s Science Mission directorate.
This work has made use of data from the European Space Agency (ESA) mission Gaia (\url{https://www.cosmos.esa.int/gaia}), processed by the Gaia Data Processing and Analysis Consortium (DPAC; \url{https://www.cosmos.esa.int/web/gaia/ dpac/consortium}). Funding for the DPAC has been provided by national institutions, in particular the institutions participating in the Gaia Multilateral Agreement. 
This research has made use of NASA's Astrophysics Data System and the NASA Exoplanet Archive operated by the California Institute of Technology, under contract with NASA under the Exoplanet Exploration Program.

\section*{Data Availability}

Parameter summaries for binaries and exoplanets in the sample, along with the entire control sample, are available as supplementary tables. Individual samples for systems are available from the corresponding author upon request.



\bibliographystyle{mnras}
\bibliography{main} 





\label{lastpage}
\end{document}